\documentclass[prb,twocolumn,amsmath,amssymb,amsfonts]{revtex4}
\usepackage[pdftex]{graphicx}
\pdfoutput=1
\usepackage{color}
\newcommand{\ud}{\mathrm{d}}

\newcommand{\ycm}{y_{cm}}

\begin{document}

\title{Carrier multiplication yields in PbS and PbSe nanocrystals
measured by transient photoluminescence}

\author{Gautham Nair}
\author{Scott M Geyer}
\author{Liang-Yi Chang}
\author{Moungi G Bawendi}
\email{mgb@mit.edu}

\affiliation{Department of Chemistry, Massachusetts Institute of Technology, 77 Massachusetts Avenue, Cambridge, Massachusetts 02139}

\begin{abstract}
We report here an assessment of carrier multiplication (CM) yields in PbSe and PbS nanocrystals (NCs) by a quantitative analysis of biexciton and exciton dynamics in transient photoluminescence decays. Interest in CM, the generation of more than one electron and hole in a semiconductor after absorption of one photon, has renewed in recent years because of reports suggesting greatly increased efficiencies in nanocrystalline materials compared to the bulk form, in which CM was otherwise too weak to be of consequence in photovoltaic energy conversion devices. In our PbSe and PbS NC samples, however, we estimate using transient photoluminescence that at most $0.25$ additional e-h pairs are generated per photon even at energies $\hbar\omega > 5E_g$, instead of the much higher values reported in the literature. We argue by comparing NC CM estimates and reported bulk values on an absolute energy basis, which we justify as appropriate on physical grounds, that the data reported thus far are inconclusive with respect to the importance of nanoscale-specific phenomena in the CM process.
\end{abstract}

%Interest in CM, the generation of more than one electron and hole in a semiconductor after absorption of one photon, has renewed in recent years because of reports suggesting greatly increased efficiencies in nanocrystalline materials compared to the bulk form, in which CM was otherwise too weak to be useful in the photovoltaic energy conversion devices where its most important practical potential could lie. 
%We argue that a comparison of NC CM estimates and reported bulk values on an absolute energy basis, which we justify as appropriate on physical grounds, indicates that the data thus far is inconclusive in determining the importance of nanoscale-specific phenomena in the CM process.
%We argue that comparing NC CM estimates and reported bulk values on an absolute energy basis, which we justify as appropriate on physical grounds, is not conclusive as to the role of nanoscale-specific phenomena in the CM process.
%We argue that comparing our data, previous literature results, and reported bulk values on an absolute energy basis, which we justify as appropriate on physical grounds, suggests that nanoscale-specific phenomena may not play a very large role in the CM process.
%A comparison of reported bulk CM values with our own data and the NC literature on an absolute energy basis, which we justify on physical grounds, argues that CM enhancement in NCs over the bulk is modest at best, given the experimental evidence to date.

\maketitle

\section{Introduction}

	The process of carrier multiplication (CM) consists of the generation of more than one electron and hole after absorption of a single photon in a semiconductor, its effectiveness determined by a rich interplay of the interactions between charge carriers, phonons and light.\cite{KanePR67} From a practical perspective, though, its chief potential as an enabler of more efficient solar spectrum harvesting in energy conversion devices has been limited by the very weak CM of bulk materials.\cite{WolfJAP98} The topic of CM has however reemerged in recent years due to reports of very strong enhancements of the CM process in nanocrystalline semiconductors.
	
	Enhanced CM was first reported for PbSe and PbS nanocrystals (NCs) by Schaller et al. \cite{SchallerPRL04} and Ellingson et al. \cite{EllingsonNL05} using the transient absorption (TA) technique. 
	Work on this material system has been extended, with one report inferring the creation of up to 7 e-h by a single high energy photon based on pump-probe data, \cite{SchallerNL06} and a study suggesting that the enhancement occurs not only for NCs in solution but also in close-packed films relevant for potential device applications. \cite{BeardNL07}
	Other material systems have also been explored, with work initially showing evidence for strong CM as well in CdSe, \cite{SchallerAPL05,SchallerJPCB03} InAs, \cite{PijpersJPCC07,SchallerNL07inas} and Si NCs. \cite{LutherNL07}
	
	Since then, there have been several reports observing little or no CM. Using a transient photoluminescence  experiment, we found no evidence for CM in CdSe NCs at energies well above previously reported thresholds. \cite{NairPRB07} More recently, Pijpers et al. have reported difficulty in reproducing their observation of CM in InAs \cite{PijpersJPCC08retract}, and a new study has reported no observable CM in InAs/CdSe/ZnSe (core/shell/shell) NCs. \cite{BenluluNL08} In addition, there remain several unresolved questions pertaining to CM in lead chalcogenide NCs. 
For instance, there are significant qualitative and quantitative differences between the Schaller et al. reports\cite{SchallerPRL04,SchallerPRL06} of strong CM following a universal trend with $\hbar\omega/E_g$ and the Ellingson et al. results\cite{EllingsonNL05} which appear to show a distinct particle size dependence and smaller yields, in some cases by factors of $2$-$3$. 
Second, considerable theoretical debate about CM in NCs remains, mostly due to a lack of information about intraband relaxation processes deep in the exciton and biexciton manifolds. \cite{AllanPRB06,AllanPRB08,ShabaevNL06, FranceschettiNL06, RupasovPRB07} 
%
%Though significant progress has been made, we have previously discussed that the essential nature of potential enhancement over the bulk is not yet well explained \cite{NairPRB07}. 
%
Recognizing this deficiency, Allan and Delerue have allowed for a wide range of intraband relaxation rates in their flexible theoretical framework, but still find the largest CM yields reported by Schaller et al.  difficult to accommodate. \cite{AllanPRB08} 
%
%Even then, allowing for a wide range of intraband relaxation rates in their flexible theoretical frameworks, Allan and Delerue have recently pointed out that the largest CM yields reported by Schaller et al.\cite{SchallerNL06} can still be challenging to accomodate. \cite{AllanPRB08} 
%
%These outstanding issues and our previous work on CdSe, showing the large discrepancy between CM yields determined by tPL and TA, suggest the need for a new assessment of CM in lead chalcogenide NCs.
Overall, these outstanding issues suggest the need for continuing the assessment of CM in lead chalcogenide NC samples.

	In this work we study carrier multiplication in PbSe and PbS nanocrystals using transient photoluminescence (tPL), a technique that more specifically informs on the e-h pair population within NCs than the pump-probe methods commonly employed. \cite{NairPRB07,SchallerJPCB06} 
	We first characterize the exciton and multiexciton PL signatures in these materials using low photon-energy excitation. We find that PbSe and PbS NCs, when adequately surface passivated, have flat exciton population dynamics over a 1ns window. At higher excitation power, strong features appear with fast $50$-$200\textrm{ps}$ decay lifetimes attributed to biexcitons. After these calibration steps, we measured tPL decays to look for evidence of CM using excitation at $3.1\textrm{eV}$, well above previously reported CM energy thresholds for the NC materials in this study. \cite{SchallerPRL04,EllingsonNL05} Although we distinctly observe a signal consistent with CM for all of our PbS and PbSe NC samples,  the CM yields we estimate, defined as the average number of \emph{additional} e-h pairs generated per absorbed photon, 
		%---------footnote 			
	\footnote{In the literature CM yields are often reported as an internal quantum efficiency (IQE), which is related to our $\ycm$ by $\textrm{IQE}=100\% (y_{cm}+1)$.}
	%---------end footnote
	reach only $\ycm\approx 25\%$ even when $\hbar\omega>5E_g$. These values are significantly lower than those of previous reports.\cite{SchallerNL06,EllingsonNL05}
	
	In the final section we explore the issue of comparing CM yields between NCs of different sizes and with the bulk. We show that if nanoscale-specific physics, such as potentially slowed intraband relaxation, are not \emph{a priori} assumed, one would expect CM yields to depend only on the incident photon energy, regardless of the particle's size. 
	%This suggests that CM yields be compared not at scaled energies $\hbar\omega/E_g$, as has been the norm in the literature, but on an absolute photon energy $\hbar\omega$ basis. 
	This suggests that CM yields be compared on an absolute photon energy basis. 
	Revisiting the literature in this framework shows that the reports on PbS and PbSe NCs to date do not uniformly suggest very large enhancements of the underlying CM physics when compared to what has been reported\cite{SmithJOSA58} for bulk PbS.
	
\section{Experimental}
\begin{figure}
\includegraphics*[scale=1.0,viewport=0 0 198 186]{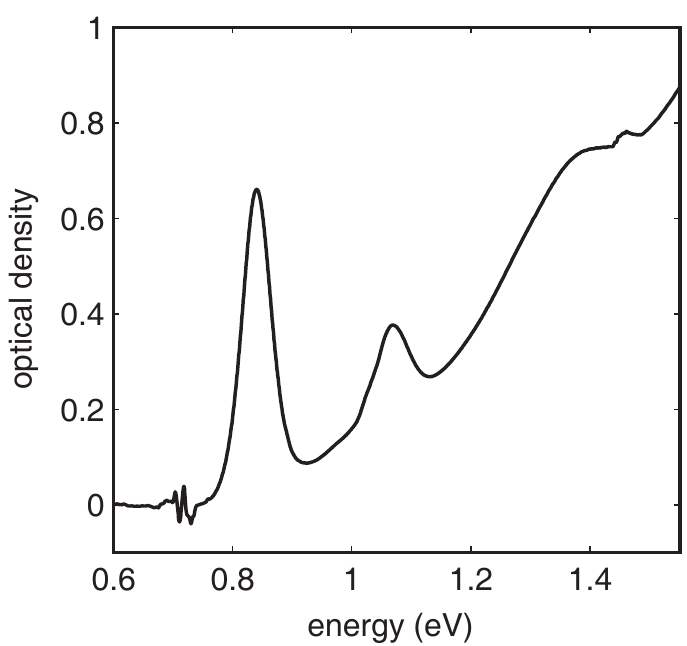}
\caption{\label{fig_absorption} Absorption spectrum of a typical PbSe NC sample used in this study. The NC bandgap $E_g=0.84 eV$ is determined as the peak of the first absorption feature.} 
\end{figure}

%\subsection{Sample preparation}
PbSe and PbS NCs were prepared by high temperature pyrolysis of Pb and Se/S precursors in an oleic acid/octadecene mixture.\cite{HinesAM03,SteckelAM03} The growth solutions were purified by a single precipitation, redispersed in hexane, and transferred to 1mm path length quartz cuvettes in a nitrogen glovebox. The resulting samples, with optical densities of $\sim 1$-$3$ at $1.55\textrm{eV}$, were sealed and taken out into air for subsequent measurements. As will be described below, some samples of larger particles (first absorption feature $<0.8\textrm{eV}$) were treated with Cd$^{2+}$ by adding a few drops of cadmium oleate to the hexane NC dispersions at room temperature. All samples were magnetically stirred during acquisitions, and PL decays under weak $1.55\textrm{eV}$ excitation were periodically monitored to check for any degradation. A typical sample's absorption spectrum is shown in Fig. \ref{fig_absorption}.

%\subsection{Experimental apparatus}
Transient photoluminescence decays of the samples were collected using a fluorescence upconversion apparatus based on an amplified Ti:sapphire system operating at 250kHz. A portion of the pulse train was passed through a BBO crystal to generate excitation sources at 3.1 eV and 1.55 eV which were separated with two dichroic mirrors and focused on the sample to spot sizes of roughly $\sim100\mu$m and $\sim50\mu$m diameter respectively. Emission was collected in a front-face geometry using off-axis parabolic mirrors and focused onto another BBO crystal. Following a variable delay, the rest of the $1.55\textrm{eV}$ pulse train was overlapped with the collected emission and the resulting sum frequency generation was separated spatially and spectrally using interference filters and a monochromator. The signal was detected with a cooled PMT and amplified using a lock-in amplifier.  
For these experiments, the pulse width was maintained relatively long by tweaking the amplifier compressor away from its optimal short-pulse configuration to avoid excess noise from continuum generation in the mixing crystal. We have nevertheless maintained a time resolution better than $\sim15$ ps as measured from the rise time of the tPL signal.
%The decays shown in this report were all acquired at a fixed wavelength corresponding to the maximum of the PL upconversion spectrum. 
Because the peaks of the exciton and multiexciton PL were not found appreciably different within our spectral resolution, all decays for a given sample were acquired at a fixed wavelength. 
%The PL upconversion technique has previously been applied by XXX to study intraband relaxation, but this is the first time MX are being seen.

\section{Results and Discussion}
\subsection{Exciton decay dynamics}
\begin{figure}
\includegraphics*[scale=1.0,viewport = 0 0 234 174]{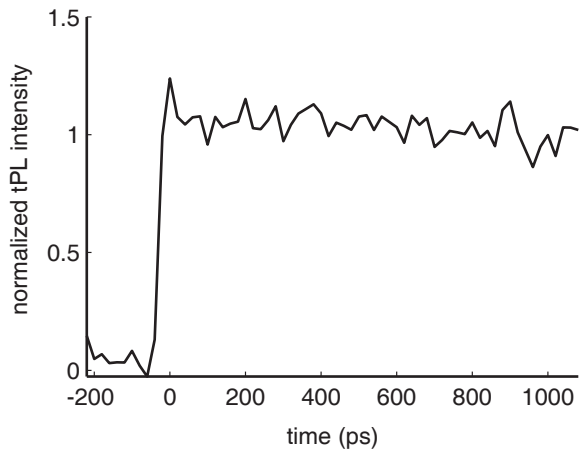}
\caption{\label{fig_xdecay} Transient PL dynamics of a sample of PbSe NCs in hexane dispersion ($E_g=0.84 eV$) under weak $1.55 eV$ excitation.} 
\end{figure}

We began by characterizing the PL dynamics of single excitons (X) in PbS and PbSe NCs using weak $1.55\textrm{eV}$ excitation. In general, samples of small and moderate sized NCs had flat PL dynamics over the full temporal range of our instrument (see Fig. \ref{fig_xdecay}). 
In contrast, as-prepared CdSe core particles almost invariably show significant sub-nanosecond dynamics, attributable to trapping by defects. \cite{NairPRB07,SchallerJPCB06} The flat decays we observe in these PbSe and PbS samples suggest good surface passivation of NCs prepared by these methods,\cite{HinesAM03,SteckelAM03} and are consistent with the very high luminescence quantum yields reported in the literature. \cite{WehrenbergJPCB02} In addition, we also measured the PL dynamics over a much longer window for one of our PbS samples using an InGaAs amplified photodiode and found a nearly single exponential fluorescence decay with a $\sim660\textrm{ns}$ lifetime, consistent with previous studies.\cite{WehrenbergJPCB02}

	The PL dynamics of larger as-prepared particles, those with $E_g < \sim 0.8 \textrm{eV}$, typically showed multiexponential X decays with large sub-nanosecond components, suggesting poor surface passivation. Moreover, these dynamics steepened irreversibly upon exposure to $3.1\textrm{eV}$ radiation. In an attempt to remove non-radiative pathways and to stabilize the particles, we chose to apply a mild Cd$^{2+}$ treatment to the NCs. Addition of cadmium oleate to hexane dispersions of $E_g=0.73\textrm{eV}$ PbS and $E_g=0.60\textrm{eV}$ PbSe NCs resulted in nearly flat single-exponential X decays and robustness to prolonged $3.1 \textrm{eV}$ irradiation, while causing no noticeable changes in the absorption spectra and emission wavelength of the samples. Our measurements also suggest that surface treatment of these samples does not have much effect on CM yields. We studied one sample of fairly large $E_g=0.68 \textrm{eV}$ PbSe NCs that did not require Cd$^{2+}$ treatment, and found that its estimated CM yields were similar to the other large NC samples that were treated with Cd$^{2+}$. We also checked the effect of the Cd$^{2+}$ treatment by applying it to NC samples that already exhibited adequate surface passivation and found no significant change of the biexciton lifetime or estimated CM yield. 
	
	We have chosen to carry out further studies only on samples that show flat tPL decays over our experimental timescale, whether as-prepared or Cd$^{2+}$-treated, because the interpretation of subsequent results is considerably simplified. A multiexponential X decay entails an inhomogeneous distribution of NC surface passivation which can then support a nontrivial distribution of multiexciton lifetimes, \cite{NairPRB07,SchallerJPCB06} complicating both the isolation of MX features in tPL decays and the quantification of the underlying exciton and multiexciton populations. The second and more serious problem was that the X decays of samples with poor surface passivation tended to change irreversibly when exposed to $3.1\textrm{eV}$ for the lengths of time necessary to obtain adequately clean data with our apparatus. For these reasons we focused only on well-passivated samples. It is conceivable that CM yields might depend on the details of the NC surface. If so, the results of this work may be difficult to generalize beyond the constraints of our particular sample preparation and selection methods.

	%It should be noted, however, that NCs optimized for certain kinds of energy conversion applications, like as sensitizers over porous transport media, would not necessarily need to meet our surface passivation criteria. 
	%We would like to note at this stage that, because of the tPL methodology, our study and CM estimation method is restricted to well passivated samples, in which no electron \emph{or} hole trapping phenomena occur within our experimental window of $10$ps-$1$ns. As discussed by Schaller et al. \cite{SchallerJPCB07} in the case of NCs, pump-probe decays can remain relatively flat even in the presence of significant trapping phenomena. Our method then necessarily implies an inherent sample, and could be the origin of some of the later
	%Samples that show flat PL decays over timescales of $\sim 1 \textrm{ns}$ as in our experiments are reasonably well passivated. 
	
\subsection{transient PL of the BX state}
\begin{figure}
\includegraphics*[scale=1.0,viewport = 0 0 204 390]{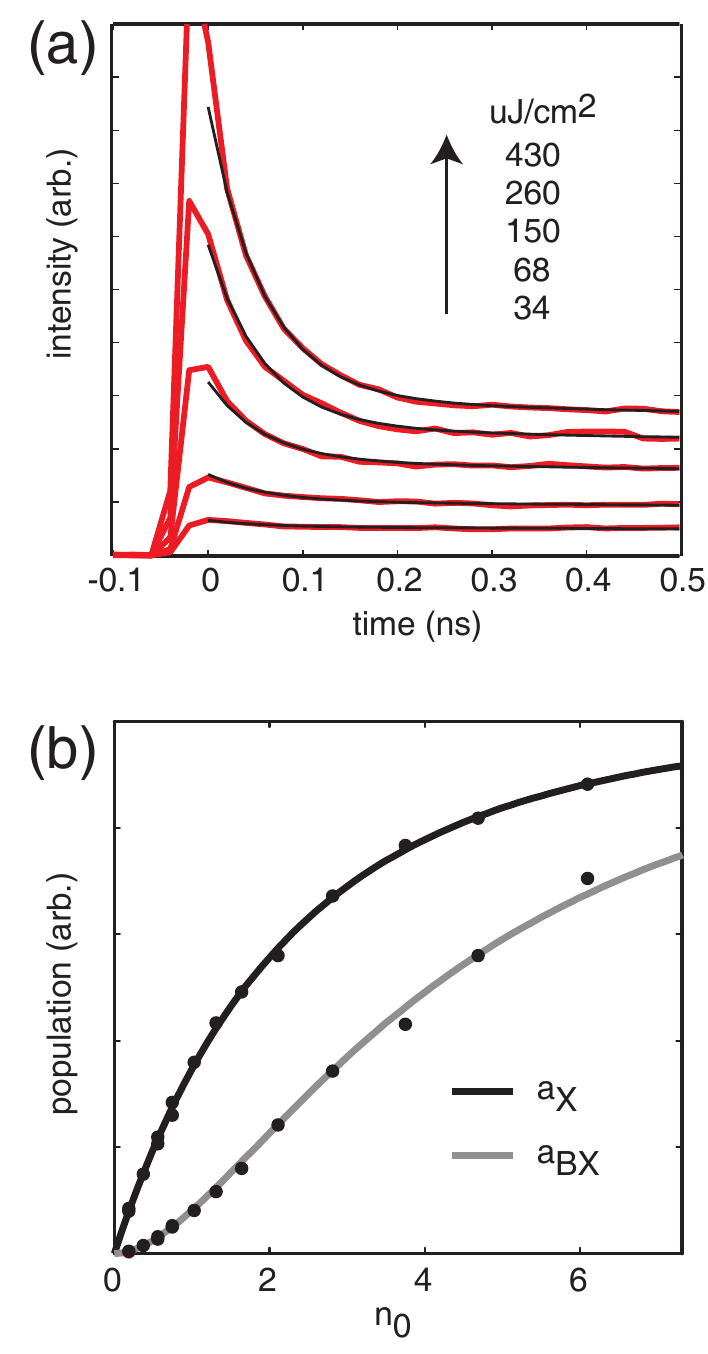}
\caption{\label{fig_800series} (Color Online). (a) Transient PL dynamics of PbSe NCs ($E_g=0.84 eV$) under increasingly strong $1.55\textrm{eV}$ excitation, showing the growth of the BX emission feature (red lines). The thin solid black lines are fits to $a_Xe^{-t/\tau_X}+a_{BX}e^{-t/\tau_{BX}}$ with $\tau_{X}>1\textrm{ns}$ and $\tau_{BX}=58\textrm{ps}$ fixed. (b) Fits of the X and BX exponential components $a_X$ and $a_{BX}$ to a population profile following poissonian photon absorption statistics for an inhomogeneous excitation beam. $n_0$ denotes the maximum average number of photons absorbed. See appendix for details.}
\end{figure}

	Strong excitation pulses can create biexcitons (BX) and further multiexcitons in NCs by sequential photon absorption. An excitation power series for our $Eg=0.84\textrm{eV}$ PbSe sample is presented in Fig. \ref{fig_800series}a, showing the growth of a large fast feature, which we attribute to the BX, on top of the single X dynamics. These decays are well described as the sum of a slow X component and a fast BX component, $a_{BX}e^{-t/\tau_{BX}}+a_Xe^{-t/\tau_X}$ with fixed lifetimes $\tau_X > 1 \textrm{ns}$ and $\tau_{BX}\approx 60 \textrm{ps}$.	Under strong excitation, additional faster components appear, attributable to emission from higher multiexcitons.
%
%------------- footnote - omitting short times from decays
\footnote{As in our previous work,\cite{NairPRB07} we delay fitting of the $1.55\textrm{eV}$-excited decays by a time $\sim \tau_{BX}/2$ to minimize unwanted interference from higher MX tPL components in the determinations of $a_X$ and $a_{BX}$.}
	The rapid $\tau_{BX}$ decay times are due to an Auger-like relaxation mechanism \cite{KlimovSCIENCE00auger} and the rates we measure are consistent with those measured by pump-probe techniques. 
%
%
%---------------footnote lifetime comparison
%
\footnote{For example, we find $\tau_{BX}\approx58\textrm{ps}$ and $\tau_{BX}\approx140\textrm{ps}$ for $E_g=0.84\textrm{eV}$ and $E_g=0.68\textrm{eV}$ PbSe NCs respectively, while Beard et al.\cite{BeardNL07} have determined $\tau_{BX}=67\textrm{ps}$ for $E_g=0.84\textrm{eV}$ and Schaller et al.\cite{SchallerNL06} report $\tau_{BX}=149\textrm{ps}$ for $E_g=0.64\textrm{eV}$.}
% Beard gets for QDs in TCE - 0.84 eV, 67 ps. I have 58 ps for 0.84eV
% Schaller gets 149 ps for PbSe at 0.64 eV. I have ~140 ps for 0.68 eV and 154 ps at 0.6 eV
%
%---------------end lifetime comparison
%
% 
% kX was 0.00014 /ps for these dots

	Our method for estimating CM yields,\cite{NairPRB07} described in the next section, relies importantly on an accurate calibration of the link between observed tPL decays and the underlying BX and X populations soon after excitation. This information can be summarized in the quantity $(a_{BX}/a_X)_{sat}$, the ratio of the sizes of the BX and X tPL decay components expected in the hypothetical case that all NCs are initially excited to the BX state, i.e. when the BX is saturated. In Fig. \ref{fig_800series}b, we fit the observed exponential components, $a_X$ and $a_{BX}$, to population profiles assuming poissonian photon absorption statistics. The power series of X and BX features are found consistent with this assumption, and we are able to estimate sample-dependent $(a_{BX}/a_X)_{sat}$ values in the range of 2.5-4. This implies that the radiative rate of the biexciton, $k_{BX}^{\textrm{rad}}$, is $\approx3.5-5$ times greater than $k_{X}^{\textrm{rad}}$. Interestingly, the numbers are similar to those observed for CdSe NCs, where we proposed that the enhanced $k_{BX}^{\textrm{rad}}$ could be due to spin substructure since the lowest X fine structure state in CdSe is known to be dark \cite{EfrosPRB96,NirmalPRL95} but the ground state BX is predicted to be bright. \cite{ShumwayPRB01} However, it has been suggested that no such spin structure is necessary to explain the long X emission lifetimes of lead chalcogenide NCs.\cite{WehrenbergJPCB02} In such a scenario, a simple accounting of all the possible electronic configurations of band-edge X and BX assuming known selection rules and thermal equilibrium gives $k_{BX}^{\textrm{rad}}=4k_{X}^{\textrm{rad}}$, which is consistent with our results. A derivation can be found in the appendix, along with an explanation of the relationship between $(a_{BX}/a_X)_{sat}$, $k_{BX}^{\textrm{rad}}$, and $k_{X}^{\textrm{rad}}$, and a description of our population profile modeling.
	
	%t might be due simply to the bulk interband dipole matrix element the dielectric constant and the photon density of states at the emission wavelength. 	
	%However, it is not known what role such spin fine structure plays in emission from lead chalcogenide NCs. It has been suggested that no such fine structure is necessary to explain the very long emission lifetimes of PbX nanocrystals, and that it might be due simply to the bulk interband dipole matrix element, the dielectric constant and the photon density of states at the emission wavelength. If so, one would expect that $k_{BX}^{rad}=2k_{X}^{rad}$. Our results would be inconsistent with such a prediction, suggesting that, like in the CdSe case, the PbX exciton radiative decay is also slowed down by dark states in its fine structure. 
\subsection{Carrier Multiplication}
\begin{figure}
\includegraphics*[scale=1.0,viewport = 6 36 222 510]{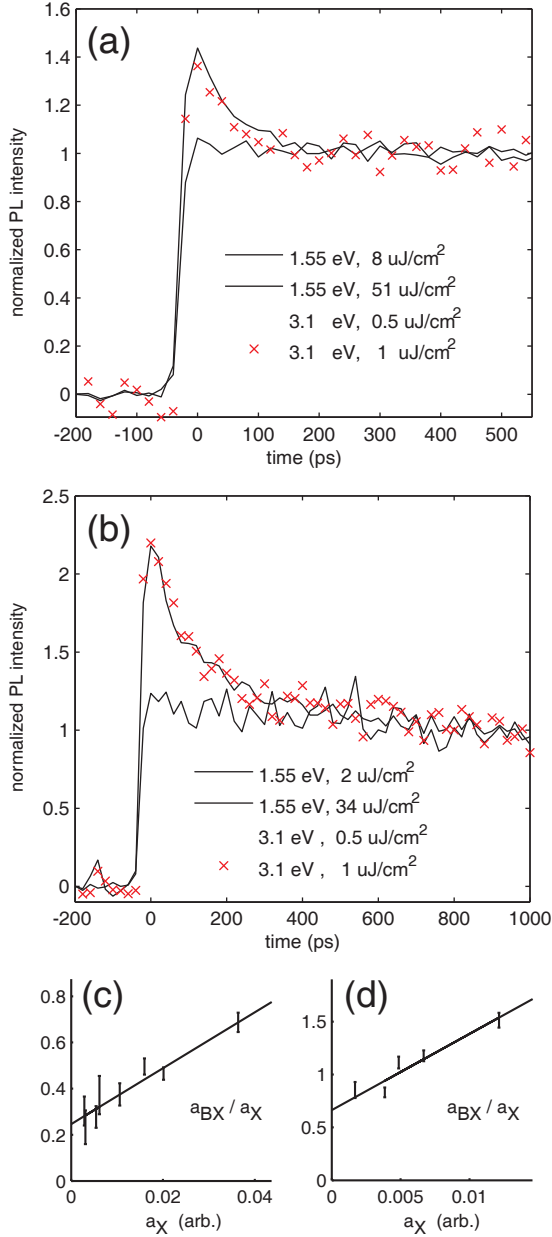}
\caption{\label{fig_400vs800} (a) Comparison of PL decays from a sample of $E_g=0.84\textrm{eV}$ PbSe NCs under $1.55 \textrm{eV}$ and $3.1 \textrm{eV}$ excitation. Even as the $3.1 \textrm{eV}$ excitation power reaches the low power limit, the decays continue to exhibit a fast component consistent with BX dynamics. (b) Same as (a) but for $E_g=0.68\textrm{eV}$ PbSe NCs. (c)-(d) Plots of $a_{BX}/a_X$ vs. $a_X$ for different weak $3.1\textrm{eV}$ excitation fluences and extrapolation to the $a_X\rightarrow 0$ ($P\rightarrow 0$) limit for the samples in (a) and (b) respectively. Dividing this extrapolated value by the $(a_{BX}/a_X)_{sat}$ determined from an independent $1.55\textrm{eV}$ power series gives CM yields $\ycm=0.09$ and $\ycm=0.23$ for the two samples at $\hbar\omega = 3.7 E_g$ and $\hbar\omega = 4.6 E_g$ respectively.}
\end{figure}

	We turn now to studying tPL decays under $3.1\textrm{eV}$ excitation. Photons of this energy are well above the CM thresholds that have been previously reported for PbS and PbSe NCs. \cite{SchallerNL06,EllingsonNL05} If carrier multiplication occurs in our samples, it would be reflected in the tPL dynamics as a residual BX or higher MX component that persists in the limit of very weak excitation, when at most one photon is absorbed per NC. Fig. \ref{fig_400vs800}a compares PL dynamics for $E_g=0.84\textrm{eV}$ PbSe  NCs under $1.55\textrm{eV}$ and $3.1\textrm{eV}$ excitation. 
	As described previously, weak $1.55\textrm{eV}$ excitation results in flat, single exponential dynamics corresponding to X decay, while at higher power the tPL traces exhibit a fast BX component as well. In contrast, even at low fluence, excitation at $3.1\textrm{eV}$ results in decays with a fast component closely following BX dynamics. Fig. \ref{fig_400vs800}c shows the $a_{BX}/a_X$ ratios obtained from a series of  measurements with varying $3.1\textrm{eV}$ excitation fluence. Our extrapolation shows that the BX-like feature persists in the zero power limit ($P\rightarrow 0$), and we thus attribute it to CM. 
The CM yield, $\ycm$, for the sample is then given by: \cite{NairPRB07}
\begin{displaymath}	y_{cm}=\left(\frac{a_{BX}}{a_X}\right)_{P\rightarrow0}/\left(\frac{a_{BX}}{a_X}\right)_{sat}
\end{displaymath}
	Because our best estimates of $\left(a_{BX}/a_X\right)_{sat}$ are in the range $2.5-4$, CM yields are smaller by a factor of $\sim3$ than the simple ratio $a_{BX}/a_X$ would suggest. For this $E_g=0.84\textrm{eV}$ sample, $\ycm\approx 9\%$ at $\hbar\omega/E_g=3.7 $. Fig. \ref{fig_400vs800}b and \ref{fig_400vs800}d display similar data for a sample of larger $E_g=0.68\textrm{eV}$ PbSe NCs. The sample exhibits a bigger fast component in the $P\rightarrow 0$ limit, and therefore a larger CM yield of $\approx 23\%$ at $\hbar\omega/E_g=4.6$. 
	%We note that although this is a sample of relatively large NCs, it was sufficiently well-behaved that Cd treatment was not necessary.
	
\begin{figure}
\includegraphics*[scale=1.0,viewport = 0 0 228 186]{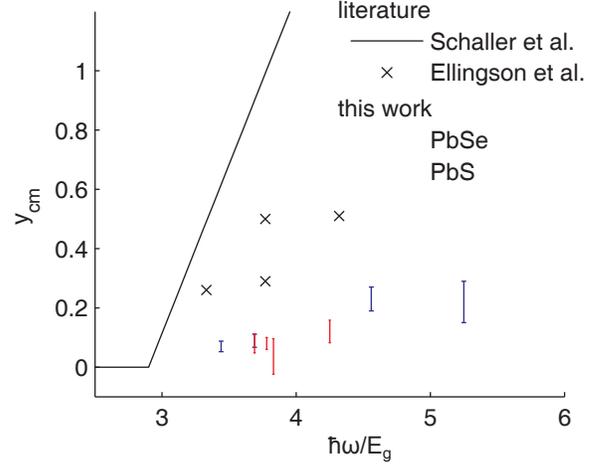}
\caption{\label{fig_pbxcmvserel} (Color online). Summary of CM yields determined in this study and comparison to literature reports on PbSe and PbS NCs at $\hbar\omega \approx 3.1 \textrm{eV}$ of Schaller et al.\cite{SchallerNL06} and Ellingson et al.\cite{EllingsonNL05} Error bars show approximate $95\%$ confidence intervals reflecting uncertainties due to noise in experimental decays.}
\end{figure}

	%Fig. \ref{fig_400vs800} shows a summary of our CM determinations for all samples used in this study, and a comparison with the literature on CM in PbSe and PbS NCs. 
	%These values are substantially below those reported in the literature.
	%To narrow the scope of the comparison, we compare only to CM determinations in the literature at $\hbar \omega = 3.1 eV$, which is the only UV excitation employed in our present study. 
	We have studied a number of PbS and PbSe NC samples in this way and find $\ycm$ always in the range of $10-25\%$ even in samples for which $\hbar \omega > 5 E_g$. As is summarized in Fig. \ref{fig_pbxcmvserel}, our CM yield estimates are significantly lower than those previously reported by other researchers for their PbS and PbSe NC samples. Schaller et al. report a universal dependence of CM yields on $\hbar \omega/E_g$, predicting CM yields of up to $2$ additional e-h pairs at $\hbar \omega = 5 E_g$. \cite{SchallerPRL04,SchallerNL06} Ellingson et al. do not observe such structure in their data, but instead report CM yields that appear to depend on particle size. \cite{EllingsonNL05} The Ellingson et al. data fall bellow the universal curve of Schaller at al. by a factor of roughly two, while our own best estimates of the CM yields are an additional factor of 2-3 smaller. It should also be noted that the findings of Schaller et al. predict not only BX formation, but also triexciton (TX) yields of 0.5 and 1 for our largest samples when excited at $3.1\textrm{eV}$. However, our data fits very well to only a BX and an X component. Any appreciable TX would have been evidenced in our measured decays since the TX decay dynamics are within our experimental time resolution and the TX emission peak is expected to be close to that of the X and BX because of the approximate 8-fold degeneracy of the lowest lead chalcogenide NC electron and hole states. 
	%Our data is therefore consistent only with $\ycm<1$ even at photon energies in excess of $5E_g$. 

	%\subsection{Robustness of CM determination}
	Since our numerical results are in disagreement with the previous reports based on TA techniques, we consider here possible sources of error in our CM estimates. In Fig. \ref{fig_pbxcmvserel} we show estimated $95\%$ confidence intervals for $(a_{BX}/a_X)_{P\rightarrow 0}$ related to noise in the experimental decays. These uncertainties in $\ycm$ are all smaller than $\pm 0.06$ and are likely unbiased. The part of our methodology most susceptible to a systematic error is the saturation ratio of the BX to X tPL components, $(a_{BX}/a_X)_{sat}$. Any multiplicative error in this quantity translates directly into a multiplicative error in the CM yield. In our study, we have estimated $(a_{BX}/a_X)_{sat}$ by fitting the sizes of X and BX decay components under $1.55\textrm{eV}$ excitation to a population profile and then assuming that this saturation ratio should apply as well to biexcitons created by a CM process. Using these $(a_{BX}/a_X)_{sat}$ values of $\approx2.5$-$4$, we have determined CM yields in the range of $10$-$25\%$. For our results to roughly match the magnitudes of CM reported by Ellingson et al. we would have had to use much smaller values $(a_{BX}/a_X)_{sat} \sim 1$, with an even further reduction to $(a_{BX}/a_X)_{sat} < 1$ required to achieve agreement with the Schaller et al. reports. 
	%The latter in particular would suggest very small BX radiative rates $k_{BX}^{\textrm{rad}}<2 k_X^{\textrm{rad}}$. 
	However, using such small $(a_{BX}^{rad}/a_X^{rad})_{sat}$ would be inconsistent with our direct observation of $a_{BX}>2a_X$ under sufficiently strong excitation conditions. We are therefore confident in our principal conclusion that the CM yields in the PbSe and PbS NC samples we have studied are significantly smaller than those previously reported for the PbX material system. 
	%We therefore regard our principal conclusion, that CM yields in the PbSe and PbS NC samples we have studied are significantly smaller than previously reported, to be quite robust.

\subsection{Comparison with bulk CM}
	In this section we seek to establish a basis for comparison of CM yields between NC samples of different sizes and with the bulk material. It has been common to compare CM yields at the same scaled energies $\hbar\omega/E_g$. This practice follows precedent from the bulk impact ionization literature and is useful when considering device applications. 
	However, aside from providing a convenient way to show data from different materials on a single plot, the physical basis for such comparisons is not obvious. 
	It may not transparently lead to answers of some basic questions, like whether or not nanoscale-specific phenomena have a large effect on CM. 
	In general, the CM yield for a material system (for example, CdSe or PbS) is expected to be determined both by particle size and the photon energy, $\ycm(r,\hbar\omega)$, which can be recast as $\ycm(E_g,\hbar\omega)$, where $E_g$ is the size-dependent bandgap. 
	Much of the existing NC CM literature infers an important role for nanoscale physics from the fact that their estimates of $\ycm$ are much larger than reports for $\ycm^{bulk}$ when compared at the same relative energy $\hbar\omega/E_g$. This assumes that without enhancement $\ycm(E_g,\hbar\omega)=\ycm^{bulk}\left[(E_g^{bulk}/E_g)\hbar\omega \right]$, or, in other words, that at a given $\hbar\omega$, NCs would exhibit only the CM that would be present in the bulk at the lower photon energy $(E_g^{bulk}/E_g)\hbar\omega$. 
	%	
	%Looking for nanoscale enhancement by comparing NCs and the bulk at the same scaled energy $\hbar\omega/E_g$ implicitly assumes that $\ycm(E_g,\hbar\omega)= \ycm^{bulk}\left[E_g^{bulk}(\hbar\omega/E_g)\right]$ in the absence of nanoscale-specific phenomena. 
	%A comparison between different sized particles or between NCs and the bulk at the same scaled energy $\hbar\omega/E_g$ implicitly makes the reference case assumption that $\ycm(E_g,\hbar\omega)=\ycm^{bulk}\left[E_g^{bulk}(\hbar\omega/E_g)\right]$. 
	%The particle size dependence of mechanisms that could enhance CM in NCs is not well understood, especially at the high excess energies where efficient CM has been reported, and the energy gap is itself another function of $r$. 
	%
	To our understanding, though, the only property of $\ycm(E_g,\hbar\omega)$ that \emph{a priori} scales with $E_g^{bulk}/E_g$ is the energy conservation requirement, $\ycm(E_g,\hbar\omega)=0$ for $\hbar\omega<2E_g$, but this does not seem sufficient to justify the assumption that $\ycm(E_g,\hbar\omega)=\ycm^{bulk}\left[(E_g^{bulk}/E_g)\hbar\omega \right]$ in general as an adequate description of CM physics in the absence of NC enhancement.
	%
	%To our understanding, though, the only property of $\ycm(E_g,\hbar\omega)$ that \emph{a priori} scales with $\hbar\omega/E_g$ is the energy conservation requirement, $\ycm(E_g,\hbar\omega)=0$ for $\hbar\omega<2E_g$, and is not a sufficient condition to imply a universal behavior $\ycm(E_g,\hbar\omega)=f(\hbar\omega/E_g)$ also for $\hbar\omega>2E_g$.
	%In fact, we argue below that $\ycm(E_g,\hbar\omega)$ should equal $\ycm^{bulk}(\hbar\omega)$ as long as energy conservation is satisfied
	%
	%
	%This would be a surprising result except for when $\hbar \omega$ is very close to $2E_g$ and energy conservation is an important constraint. 
	%However, all reports for PbSe and PbS show thresholds for large CM around $3E_g$, well over the $2E_g$ lower bound. At these high energies, the particle size dependence of mechanisms that could enhance CM in NCs is not well understood, and the energy gap is itself another function of $r$. On these grounds alone,  $\hbar \omega / E_g(r)$.
		
\begin{figure}
\includegraphics*[scale=1.0,viewport = 6 6 234 174]{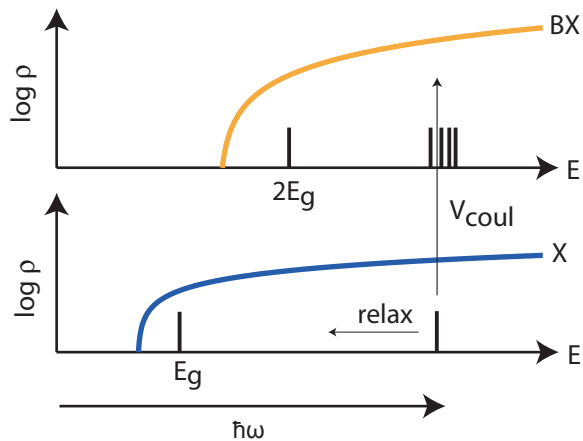}
\caption{\label{fig_xbxdosdiagram} (Color online). Diagram of relevant features and processes for bulk and NC carrier multiplication. The smooth curves are schematics of the bulk 1e1h and 2e2h densities of state, corresponding to X and BX states in an NC. Shown for the case of a NC are the lowest X and BX states at $E_g$ ($>E_g^{\textrm{bulk}}$) and $2E_g$ and a representative X state formed immediately after absorption of a high energy photon ($\hbar\omega \gg 2E_g$) subject to subsequent intraband relaxation down the X manifold or coulomb coupling to the BX states.}
\end{figure}
		
		%In creating a framework for comparison, it is important to keep in mind that the central question underlying the NC studies is whether or not there are physical processes that enhance CM compared to the bulk. 
		To construct a more appropriate reference for comparison with NC results, we consider bulk material physics and explore how $\ycm(r,\hbar\omega)$ behaves if all phenomena exclusive to the nanoscale are neglected. 
		In the bulk limit it is intuitively clear that $\ycm(r,\hbar\omega)$ is independent of $r$. The competing processes of intraband relaxation and impact ionization have the same rates for crystals of, say, $1\mu m$ and $0.5 \mu m$, resulting in the same CM efficiency. To understand why the impact ionization rate remains constant one can start from the first-order perturbation theory formulation:
		\begin{displaymath}
		k_{1e1h\rightarrow 2e2h}=\frac{2\pi}{\hbar}\overline{|\langle 2e2h | V_{\textrm{coul}} | 1e1h \rangle|^2} \rho_{2e2h}(E)
		\end{displaymath}
	Where $V_{\textrm{coul}}$ is the coulomb interaction and $\rho_{2e2h}(E)$ is the density of two electron, two hole states (corresponding to a BX) at the energy $E$ of the initial one electron, one hole configuration (which corresponds to X in an NC). A reduction in volume has two effects. First, the average coulomb coupling is enhanced, with $\overline{|\langle 2e2h | V_{\textrm{coul}} | 1e1h \rangle|^2} \propto V^{-4}$. 
%
%-----------------footnote Vcoul selection rules ------------
%
\footnote{The steep volume dependence $\overline{|\langle 2e2h | V_{\textrm{coul}} | 1e1h \rangle|^2} \propto V^{-4}$ might appear surprising at first. It should be kept in mind that this square matrix element is averaged over \emph{all} 2e2h configurations of nearby energy. However, for $\langle 2e2h | V_{\textrm{coul}} | 1e1h \rangle\neq0$, conservation of momentum and spin must be satisfied and one of either the initial electron or hole must not change state. The proportion of final 2e2h states that violate these conditions and thus have $\langle 2e2h | V_{\textrm{coul}} | 1e1h \rangle=0$ increases with volume, ultimately resulting in a stronger volume scaling of $\overline{|\langle 2e2h | V_{\textrm{coul}} | 1e1h \rangle|^2}$ than would be expected from averaging only the non-zero terms. }	
%
%-----------------end footnote Vcoul selection rules-----------
%
	However, this is fully balanced by the reduction in average density of states (DOS), since $\rho_{2e2h}(E)\propto V^4$. If no new physics are introduced, this process of shrinking the bulk can be continued into the nanoscale with the important conclusion that for $\hbar\omega$ above the $2E_g(r)$ energy conserving threshold,  $\ycm(r,\hbar\omega)=\ycm^{bulk}(\hbar\omega)$. 
	It must be kept in mind that even though the spacing between energy levels is certainly larger in NCs, the DOS averaged over sufficiently wide intervals is the same as in the bulk, with volume scalings $\rho_{X}(E)\propto V^{2}$ and $\rho_{BX}(E)\propto V^{4}$. 
	We should only expect a divergence from $\ycm(E_g,\hbar\omega)=\ycm^{bulk}(\hbar\omega)$ if new physics appear in the nanoscale that have a strong influence on the CM process. 
	%This is in contrast to the commonly used $\ycm(E_g,\hbar\omega)=\ycm^{bulk}\left[ \hbar\omega (E_g^{bulk}/E_g) \right]$ reference.	

	This suggests that comparisons between NC and the bulk should be made on an absolute photon energy basis as long as $\hbar\omega$ is well above the energy-conserving limit. Then the difference $\ycm(E_g,\hbar\omega)-\ycm^{bulk}(\hbar\omega)$ would be attributable specifically to nanoscale phenomena.
	%Deviations could be attributed to essentially nanoscale phenomena, but the comparison of CM yields, either between NCs of different sizes, and more importantly with the bulk, needs to be made at the same \emph{absolute} energy, $\hbar\omega$, irrespective of $E_g(r)$ (as long as the photon energy is well above the energy-conserving limit). 
	In contrast, the usual literature comparison at fixed $\hbar\omega/E_g$ can significantly exaggerate enhancement over the bulk simply because $E_g(r)>E_g^{\textrm{bulk}}$, so that, for instance, even without novel NC physics, PbSe and PbS NCs with $E_g(r)>2E_g^{\textrm{bulk}}$ will appear to show at least a two-fold CM threshold reduction. It is noted that from a practical perspective, bulk-like CM in NCs does indeed present a real advantage because the extra carriers can be extracted at a higher voltage difference $E_g(r)$. Also, a $\hbar\omega/E_g$ basis is useful in comparing a sample's actual CM to the maximum possible imposed by energy conservation. However, it is not obvious that a comparison of how near two \emph{different} samples are to their separate energy-conserving limits can usefully inform on differences in their underlying physics. For that, we argue that the absolute photon energy basis appears to be more appropriate.
	
\begin{figure}
\includegraphics*[scale=1.0,viewport = 6 0 222 366]{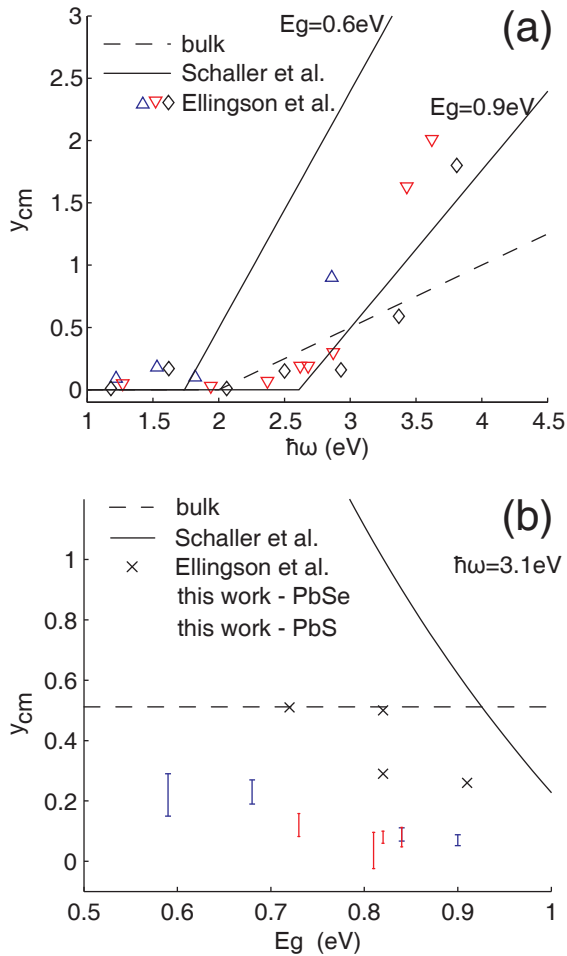}
\caption{\label{fig_cmvsE} (Color online) (a) Summary of CM reports in the literature for PbSe and PbS NCs by Schaller et al.\cite{SchallerNL06} and Ellingson et al.\cite{EllingsonNL05} compared on an absolute excitation energy scale to bulk PbS values reported by Smith and Dutton.\cite{SmithJOSA58} For the Ellingson et al. data, filled circles or open shapes indicate data taken with intraband or band-edge probe beams, while the colors blue, green, red, and black denote $E_g=0.72\textrm{eV}$, $0.82\textrm{eV}$, $0.91\textrm{eV}$ PbSe and $0.85\textrm{eV}$ PbS NC samples. \cite{EllingsonNL05} Two curves are displayed based on the Schaller et al. universal $\hbar\omega/E_g$ dependence for $E_g=0.6 \textrm{eV}$ and $E_g=0.9\textrm{eV}$. (b) Plot of CM yields against the size dependent bandgap $E_g$ for $\hbar\omega=3.1\textrm{eV}$. The dashed line is the CM yield reported for the bulk at this excitation energy.}
\end{figure}
	
	In light of these considerations, we show in Fig. \ref{fig_cmvsE}a a summary of literature data on PbS and PbSe NCs \cite{EllingsonNL05,SchallerNL06} replotted on an absolute energy axis along with values of CM yields in bulk PbS films reported by Smith and Dutton. \cite{SmithJOSA58} 
	These authors studied the photoconductivity of commercial PbS films and found an increase in photocurrent response at shorter wavelengths which they attributed to a CM process, emerging from a threshold $\hbar\omega\approx 2\textrm{eV}$ and rising approximately linearly to $\ycm^{\textrm{bulk}}\approx2$ at $\hbar\omega=6\textrm{eV}$. It should be kept in mind that there are numerous potential sources of error in this bulk CM determination, some of which we detail later, but it is nevertheless interesting to note that the CM yields for NCs reported in the literature appear only modestly enhanced over these bulk values. Except in the case of very large (small $E_g$) NCs following the universal curve of Schaller et al., CM yields are within a factor of $\approx2$ of the bulk report, and exhibit a similar CM energy threshold between $2$-$3\textrm{eV}$. 
		Figure \ref{fig_cmvsE}b shows our own estimates compared to the NC and bulk literature data at $\hbar\omega=3.1eV$ plotted against $E_g(r)$. Our results are below the bulk CM reported value, those of Ellingson et al. \cite{EllingsonNL05} appear consistent with it, and the Schaller et al.\cite{SchallerNL06} results fall well above for larger NCs.
		
	 Reaching a robust conclusion at this stage on the relative strengths of CM in bulk and NC forms is difficult because of potential uncertainties in the bulk values reported Smith and Dutton. First, the authors did not present a characterization of the commercial PbS films studied, and it is possible that significant oxidation may have taken place since no protective coating was used. \cite{SmithJOSA58} This is important since exposure to $\textrm{O}_2$ is known to cause significant changes in bulk PbS photoconductivity. \cite{BubeBOOK92} Second, the reported yields are very sensitive to any systematic errors in determining the number of photons absorbed by the film. The third complication is the possible variation of photoconductive gain with $\hbar\omega$. For example, at blue wavelengths carriers are generated on average closer to the film surface, where the greatest concentration of trap states are expected to reside. Moreover, it is difficult to say \emph{a priori} whether the gain would increase or decrease. These considerations highlight the need for a careful determination of CM in bulk films of PbS and PbSe before a definitive comparison with NCs can be made.  With the data at present it is difficult to conclude that nanoscale phenomena are responsible for strong CM enhancement, as we have discussed in the previous paragraph and in Fig. \ref{fig_cmvsE}.  

	Given the possibility that CM in NCs might follow largely bulk-like physics, it is interesting to examine what is known about the NC-specific physical mechanisms that could affect the multiplication process. 
	%The two most prominent examples are potential Coulomb coupling enhancement in excess of the bulk volume scaling and slowdown in intraband relaxation for example due to a phonon-bottleneck mechanism. It is also possible that the discrete structure of NC states might render a treatment in terms of an average DOS inadequate and perhaps induce more exotic effects like a breakdown in the Fermi Golden Rule approach if the interaction between individual X-BX states nears the strong-coupling regime. 
	%We argue that none of these effects is known to be strong enough that large deviations from $\ycm(E_g,\hbar\omega)=\ycm^{bulk}(\hbar\omega)$ should \emph{a priori} be expected. 
	The most commonly cited rationalizations of CM enhancement in NCs are the possibility of strong coulomb interaction and slow intraband relaxation. \cite{NozikARPC01} It could be argued, for example, that coulomb couplings are significantly enhanced in the nanoscale based on the much faster Auger relaxation rates of band-edge multiexcitons compared to the bulk. 
	%However, this enhancement at the band edge does not imply any similar enhancement at the high kinetic energies at which CM has been reported to take place. 
	This enhancement of Auger rates at the band edge is thought to be due to a relaxation of momentum conservation requirements brought about by the finite nature and abrupt surface of NCs.\cite{WangPRL03,EfrosBOOK03} However, because momentum conservation is not a limiting constraint on impact ionization in the bulk at high excess kinetic energies,\cite{AllanPRB06} and it is not clear that the nanocrystalline form should still exhibit significant enhancement. Calculations by Allan and Delerue suggest that $k_{1e1h\rightarrow 2e2h}$ is if anything smaller in PbSe NCs than for the bulk. \cite{AllanPRB06}
	
	Similarly, there is still no evidence of a phonon bottleneck for intraband relaxation at high electron and hole kinetic energies. Due to practical considerations relating to experimental time resolution, most studies on NCs have focused only on relaxation from some of the lowest excited states to the band edge. \cite{WehrenbergJPCB02,HarboldPRB05,BonatiPRB07pbse} Even then, they find very fast picosecond relaxation times. Moreover, at the high excess kinetic energies required for CM, the X manifold is much denser and it seems less likely that a phonon bottleneck effect could play a very large role. 
	
	The remaining potential nanoscale CM enhancement mechanisms have to do with the discrete state structure. 
	%The expression $\ycm(E_g,\hbar\omega)=\ycm^{bulk}(\hbar\omega)$  requires that a continuous bulk-like DOS can approximate the relevant physics of the BX manifold. 
	Certainly, the discrete nature of states in a NC is critical near the energy conservation threshold, as no CM can occur when $\hbar\omega<2E_g(r)$ even though the bulk 2e-2h DOS is finite. However, if we restrict our attention to $\hbar \omega$ well over $2E_g(r)$, as has been the case when large CM yields have been reported, \cite{EllingsonNL05,SchallerNL06} it is plausible that the BX manifold is sufficiently dense that bulk-like behavior could result. Further, even if there were deviations, we would not expect them to be monotonic in either $E_g(r)$ or $\hbar \omega$. Finally, it is possible that there could be strong coupling between X and BX,\cite{ShabaevNL06} but not enough is known about phase and population relaxation mechanisms of carriers with high kinetic energies to conclude that such effects would be important for CM.
	
	All these arguments above should not be taken as proof or justification that $\ycm(E_g,\hbar\omega)= \ycm^{bulk}(\hbar\omega)$ for NCs, but simply to show that such a conclusion would not be inconsistent with what is experimentally known about NC photophysics. Too little is understood about the physics of highly energetic carriers in NCs to make strong \emph{a priori} predictions of the role of nanoscale phenomena in CM.
	
\section{Conclusions}
The principal experimental conclusion of this work is that CM yields in our PbSe and PbS NC samples estimated by transient photoluminescence are well below the values that have been reported in the literature for PbSe and PbS NCs using transient pump-probe techniques.\cite{SchallerNL06,EllingsonNL05} It should be noted that these previous reports themselves show significant numerical disagreement between each other even though they employ nominally similar methods. 
%We have not sought in the main body of this paper to suggest ways to reconcile these differences. 
In broad terms, the variation between the reports of Schaller et al., Ellingson et al., and our own must ultimately stem from either systematic differences in data acquisition procedures, variation in the way CM is determined from observed decays, or actual sample-to-sample differences of the CM efficiency. The fact that Ellingson et al. and Schaller et al. use nearly equivalent methods for estimating CM but find conflicting results suggests that their samples are inherently different, or that the two groups handled these samples differently during the course of their experiments. In our own work there is a possibility of systematic error related to the calibration method we use, but we have argued above that this alone cannot readily account for the contrast with the existing literature. The answer may yet lie in sample-to-sample CM variation, and if so would suggest that CM in NCs is strongly affected by defects or surface ligand type and coverage. 

The second effort of this work has been to establish a basis for comparing CM efficiencies in NCs and the bulk that more clearly isolates the effects of changes in underlying physics.
We have argued that an absolute photon energy basis is more appropriate than the usual $\hbar\omega/E_g$ approach for this purpose, and  by comparison to values reported for the bulk, we found that the CM yields reported for NCs do not immediately suggest a very large role for nanoscale-specific phenomena.
 Because these bulk values themselves could be beset by large errors, it is difficult to reach a definite conclusion. A modern, robust, assessment of CM in bulk PbS and PbSe will be necessary for this to be possible. Similarly, understanding the variation in the NC CM literature will require applying multiple experimental methodologies to identical NC samples or, more importantly, the development of new spectroscopic techniques that are more specifically taylored to multiexciton detection than the population-dynamics based measurements in use at this time. A clear picture of the CM process in the transition from the bulk to the nanoscale will have to wait for experimental efforts on these two fronts.

\acknowledgments{This work was supported in part by the Department of Energy (DE-FG02-07ER46454), the NSF MRSEC program (NSF-DMR-0213282) at MIT making use of its Shared Experimental Facilities, the Harrison Spectroscopy Laboratory (NSF-CHE-011370), and the NSF-NIRT program (NSF-CHE-0507147). The authors would also like to thank A. Dorn for experimental assistance, and K. Gundogdu and R. Ellingson for useful advice.}

	%The comparison with the bulk is further complicated by large uncertainties in bulk CM values. Several earlier determinations of CM efficiencies in bulk PbS and PbSe have produced very different results than those of Smith and Dutton. These other studies show much larger CM with thresholds below < 2 eV. Given the experimental difficulties of carrying out such photocurrent quantum efficiency measurements with high accuracy, we can only conclude that our results are simply not inconsistent with bulk CM. A more careful comparison would require a modern, accurate determination of CM in bulk films of PbS and PbSe, which is not available at the present.

	\appendix
	
	\section{Population modeling}
	%\section{Determination of $(a_{BX}/a_X)_{sat}$}
	As has been described in detail in our previous work on CdSe NCs,\cite{NairPRB07} the exponential components in a tPL decay can be related to the MX and X populations immediately following excitation through the following approximate expressions:
	\begin{eqnarray}
	a_X& \propto & k_X^{rad}p_{1} \nonumber \\
	a_{BX}& \propto & \left( k_{BX}^{rad}-k_X^{rad} \right) p_{2} \nonumber
	\end{eqnarray}
	 where $p_{1}$ and $p_{2}$ are the population of NCs that start with at least an exciton or a biexciton respectively at time 0. These populations are given by $p_{m}=\sum_{k\geq m} I_k$, where the population of the $k$-th multiexciton state, $I_k$, is determined by poisson statistics, taking into account excitation beam inhomogeneity and position dependent collection efficiency:
	 \begin{displaymath}
	 I_k=\int \phi(x) \frac{n(x)^k}{k!} e^{-n(x)} \ud^3 x \qquad n(x)=\sigma j(x)
	 \end{displaymath}
	where $j(x)$ and $\phi(x)$ are the photon flux and collection efficiency at position $x$, and $\sigma$ is the absorption cross section. $n(x)$ is the average number of photons absorbed per pulse by an NC located at $x$. If $j(x)$, $\phi(x)$ and $\sigma$ were known, it would be possible to compute the $I_k$ up to a common proportionality constant and obtain, by comparison with experiment, the saturation ratio $(a_{BX}/a_X)_{sat}$, the value of $a_{BX}/a_X$ when $p_2=p_1$. However, both $j(x)$ and especially $\phi(x)$ are difficult to determine accurately in our apparatus. Instead, we exploit the fact that the shape of $p_1$ as a function of excitation power fully determines the shape of $p_2$. To see this, we note that during any of our experimental power series $n(x)$ only changes in magnitude while retaining its shape. Setting $n(x)=n_0 h(x)$, where $h(x)$ is a fixed shape and $n_0$ is a constant parameter, one can show that: 
	\begin{displaymath}
	p_2(n_0)=p_1(n_0)-n_0\frac{\partial p_1}{\partial n_0}
	\end{displaymath}
	In the above, $n_0$ can be replaced with any quantity proportional to it, such as average excitation power, so knowledge of the absorption cross section is not required. Therefore, if one finds any $h(x)$ and $\phi(x)$ so that the calculated $p_1(n_0)$ closely fits the shape of the observed $a_X$ excitation series, then the $p_2(n_0)$ calculated with the same $h(x)$ and $\phi(x)$ will be proportional to $a_{BX}$. The results of this procedure, shown in Fig. \ref{fig_800series}b, demonstrate that the $a_{BX}$ we observe match very well the trend we independently predict from the $a_X$ evolution, further supporting our assignment of this fast component in the tPL to the biexciton. Our estimate of $(a_{BX}/a_X)_{sat}$ is then  obtained as the ratio of the proportionality constants relating $a_X$ to $p_1$ and $a_{BX}$ to $p_2$.  We find saturation values $(a_{BX}/a_X)_{sat}$ of 2.5-4 using this method. Since $(a_{BX}/a_X)_{sat}=k_{BX}^{rad}/k_X^{rad}-1$, the corresponding values of $k_{BX}^{rad}/k_X^{rad}$ are in the range 3.5-5.
	 
	 	\section{BX and X radiative rates}
	 We present here a calculation of $k_{BX}^{rad}/k_{X}^{rad}$ for a simple model of the lead chalcogenide ground state. The $1S_e$ and $1S_h$ states in lead chalcogenide are eight-fold degenerate. There are four equivalent valleys in the band structure and two-fold spin degeneracy.  The possible X electronic configurations can be labeled $i_em_h$, and the BX configurations  $i_ej_em_hn_h$, where $i,j,m,n\in1\ldots8$. Because total momentum and spin must be conserved during an optical interaction, only the recombination of an electron and hole with the same $k$ and same spin is allowed. Assuming that particle momentum and spin remain good quantum numbers, each electron state is connected by a dipole transition to exactly one of the eight hole states. By symmetry, these transition dipole moments all have the same magnitude $|\mu|$. We can then calculate the radiative square transition dipole of each X and BX microstate. In the case of X, there are 8 configurations of type $1_e1_h$ with $k_{rad}=\mu^2$, and $8\cdot7$ of type $1_e2_h$ with $k_{rad}=0$. Similarly, for the BX, there are $\binom{8}{2}$ configurations like $1_e2_e1_h2_h$ with $k_{rad}=\mu^2+\mu^2$, $8\cdot7\cdot6$ configurations of type $1_e2_e1_h3_h$ with $k_{rad}=\mu^2$, and $\binom{8}{2}\binom{6}{2}$ dark $1_e2_e3_h4_h$-type states. Taking the thermal average, one finds $k_{X}^{rad}=\mu^2/8$ and $k_{BX}^{rad}=\mu^2/2$, and therefore $k_{BX}^{rad}=4k_{X}^{rad}$. This result should remain approximately valid even in the presence of perturbations that mix states with different quantum numbers or couple the electrons and holes, as long as the width of the resulting energy fine structure is sufficiently smaller than the available thermal energy, $kT$.
	 	 
%\bibliography{pbxCMbiblio}

\end{document}